\documentclass[a4paper,10pt]{article}
\usepackage{a4wide}
\usepackage[]{hyperref}
\usepackage[centertags]{amsmath}
\usepackage{amssymb}
\usepackage{latexsym}
\usepackage[dvips]{graphicx}
\begin{document}

\title{Optical properties and one-particle spectral function\\  in non-ideal plasmas}

\author{Carsten Fortmann\footnote{Corresponding
     author: e-mail: {\sf carsten.fortmann@uni-rostock.de}, 
     Phone: +49\,381\,498\,6943,
     Fax: +49\,381\,498\,6942}, G. R\"opke, A. Wierling\\
     \texttt{Institut f\"ur Physik, Universit\"at Rostock,
     18051 Rostock, Germany} }

\maketitle                   
\begin{abstract}
 
A basic concept to calculate physical features of non-ideal
plasmas, such as optical properties, is the spectral function which is
linked to the self-energy. We calculate the spectral function for a
non-relativistic hydrogen plasma in $GW$-approximation. In order to go beyond
$GW$ approximation, we include self-energy and vertex correction
to the polarization function in lowest order. 
Partial compensation is observed. 
The relation of our approach to $GW$ and $GW\Gamma$ calculations in
other fields, such as the band-structure calculations in semiconductor
physics, is discussed.
From the spectral function we derive the absorption coefficient due to
inverse bremsstrahlung via the polarization function. As a result, a
significant reduction of the absorption as compared to the
Bethe-Heitler formula for bremsstrahlung is obtained.
\end{abstract}

\section{Introduction}
\label{sec:intro}
%
%
Spectroscopy can serve as a versatile tool to infer properties of
dense and hot plasmas from the emitted
radiation\cite{Griem,Hutchinson}. In non-ideal plasmas, where the
coupling parameter can exceed unity, properties often deviate
significantly from their form in ideal plasmas due to the importance
of interaction effects \cite{Proc}. With the advent of femtosecond
laser pulses \cite{laser}, it has become possible to produce non-ideal
plasmas with table-top systems. In particular, it is nowadays possible
to create conditions similar to those in the center of astrophysical
objects such as the sun or giant planets.  Effects such as dynamical
screening, dissolution of bound states, Pauli-blocking, and the
importance of collisions have been observed \cite{Proc}.  Taking
proper account of interaction effects is a challenge to any
theoretical description of non-ideal plasmas.

%
%
Many-body perturbation theory presents a toolbox to determine various
properties of non-ideal plasmas \cite{Fetter,Mahan,KKER}. 
The use of Green's
function techniques allows for a systematic and intuitive consideration
of many-body effects. In particular, one-particle properties can be 
obtained from the one-particle Green's function or its spectral 
representation, the one-particle spectral function. A number of important
mechanisms in dense plasmas such as dynamical screening of the Coulomb
interaction can be described by partial summation of certain diagrams.
A particularly successful concept is the quasi-particle picture 
\cite{Fetter,Mahan,KKER}. However, with increasing coupling the 
quasi-particle pictures breaks down. This shows up in a broadening of the
spectral function. To go beyond the quasi-particle picture in a consistent
way poses serious problems of self-consistency. Vertex and self-energy
corrections have to be taken into account on the same footing to obey
sum rules and other exact known properties. Using Ward identities \cite{Ward}
or the Kadanoff-Baym scheme of the conserving vertex \cite{BK61} enables one
to construct consistent sets of diagrams, alas the resulting 
integral-equations are complicated to solve. 
%
%

The $GW$ approximation is a particularly scheme for the self-energy approximating
it by a product of the Green's function $G$ and an effective interaction $W$ \cite{A}.
Being introduced by Hedin in 1965 \cite{H65}, it has a long history of
applications, which is reviewed in Ref. \cite{Mahan94} and 
\cite{AG98}. However, the
full self-consistency implied in Hedin's original proposal was not carried
out so far. Instead, the ideal Green's function or a quasi-particle picture 
was used. In this manner, the $GW$ approximation was successfully applied to 
determine shifts in the ionization potential, approximations for the equation
of state, and an effective interaction to describe bound states in a
medium \cite{KKER}. Also, the band structure of different types of materials,
like semiconductors \cite{Godby}, alkali \cite{Northrup}, and 
transition metals \cite{Aryasetiawan}
was determined. Due to the efficiency of modern computer technology, it has
recently become feasible to address the self-consistency implied 
in Hedin's scheme to some extent, see e.g. \cite{S96,HB98,SG98,T01}. 
Some of these calculations improve solely on self-energy corrections, 
while others take into account vertex contributions as well. It is
customary to call the later $GW\Gamma$ approximations.
In this paper, we perform calculations for the optical properties of a hydrogen plasma at solar core conditions, as an important example for
astrophysical plasmas. Typical parameters of the solar core plasma are temperatures of about $T=100\,\mathrm{Ry}\simeq1360\,\mathrm{eV}/k_\mathrm{B}$ and
particle densities reaching $n\simeq 7\cdot 10^{24}\,\mathrm{cm}^{-3}$. These parameters justify the model of a classical (non-degenerate) and 
weakly coupled plasma,  being characterized by values of $\theta=k_\mathrm{B}T/E_\mathrm{F}\simeq 10$ for the degeneracy parameter and  
$\Gamma=(Ze^2/4\pi\epsilon_0)/(3/4\pi n)^{1/3}k_\mathrm{B}T\simeq 0.03$ for the
coupling parameter.

Central to the description of optical properties of plasmas is the
dielectric function $\epsilon(k,\omega)$ \cite{Griem}. Within the frame of the
approach presented here, it can be obtained from the polarization function, 
which is a member of Hedin's set of equations. In an earlier paper \cite{Fortmann}, 
the suppression of the bremsstrahlung cross section due to 
successive scattering
has been treated by a taking into account self-energy and vertex corrections.
However, these corrections were only considered within a one-loop
approximation. In this way, dynamical screening effects were neglected.
It is the objective of this communication to present results with an
improved one-particle spectral function as obtained with the
$GW\Gamma$ approximation. 

In Sec. \ref{sec:GWGamma}, we will review the $GW$ approximation
making use of Hedin's equations. Sec. \ref{sec:solar} presents an 
illustrative example for the $GW^0$ approximation. Implications for
the absorption coefficient are discussed in Sec. \ref{sec:absorption}.
A discussion and conclusions will be given in Sec. \ref{sec:conclude}.

If not otherwise indicated, we apply the Rydberg system of units where $\hbar=k_\mathrm{B}=1\,,e^2=2\,,\epsilon_0=1/4\pi\,,\ \mbox{and}\ m_\mathrm{e}=1/2$. 
\section{The $GW\Gamma$ approximation}
\label{sec:GWGamma}

A convenient starting point of our approach are Hedin's equations \cite{HL69}.
It is a closed set of equations relating the full Green's function
$G$, the non-interacting Green's function $G_0$, the self-energy $\Sigma$,
the dynamically screened interaction $W_{ab}$ \cite{A},
the polarization function $\Pi$, and the vertex function $\Gamma$.
In detail, the full Green's function is given by 
\begin{eqnarray}
\label{eq:Dyson}
G_a(12) & = & G_{a,0}(12) \,+\, \int\!d(34)\, G_{a,0}(13)\,\Sigma_a(34)\,
G_a(42)\,\,.
\end{eqnarray}
Here and in the following, the shorthand notation $(1)\equiv(\boldsymbol{r_1},\tau_1,\boldsymbol{\sigma}_1\ldots)$ 
for spatial variables $\boldsymbol{r}$, imaginary times $\tau$ as well as
quantum numbers such as spin is used.
The self-energy $\Sigma_a(12)$ is obtained from the dynamically screened interaction
and the vertex function according to
\begin{eqnarray}
\label{eq:sigma}
\Sigma_a(12) & = & 
i \int\! d(34)\,G_a(13)\, W_{aa} (41)\,\Gamma_{a}(32,4)\,\,\,.
\end{eqnarray}
The dynamical screened interaction is given via
\begin{eqnarray}
W_{ab} (12) & = &
V_{ab} (12)\,+\,\sum_{c} \int\!d(34) V_{ac}(13)\,\Pi_{cc}(34) W_{cb}(42) 
\end{eqnarray}
by the polarization function
\begin{eqnarray}
\label{eq:polarization}
\Pi_{aa}(12) & = &
\int\!d(34) G_a(13) G_a(41) \Gamma_{a}(34,2)\,\,\,.
\end{eqnarray}
Finally, the vertex function obeys a Bethe-Salpeter like equation
\begin{eqnarray}
\label{eq:vertex}
\Gamma_{a}(12,3) & = & 
\delta(12)\,\delta(13)\,+\,\sum_{b} \int\!d(4567)\,
\frac{\delta \Sigma_a(12)}{
\delta G_b(45)}\, G_b(46) \,G_b(75)\,\Gamma_b(67,3)\,\,\,.
\end{eqnarray}
The set of equations represents a perturbation expansion in terms of the
screened interaction $W_{ab}$ and is expected to show better convergence
properties compared to an expansion in
the inter-particle interaction $V_{ab}$.

By truncating the set of equations~(\ref{eq:Dyson})-(\ref{eq:vertex})
on a certain level, various approximations for these quantities can be
defined. A particular simple approximation is
the $G_0 W^0$ approximation \cite{H65,HL69}. It is obtained by taking the
bare vertex $\Gamma(12,3)=\delta(12) \delta(13)$ and inserting non-interacting
Green's functions into the expression for the polarization
function. This leads to a self-energy given by $\Sigma_{a}^{GW}(12)=i G_a(12)
W_{aa}(21) $. Next, the Green's functions in this expression as well as in
the screened interaction $W$ are taken as the free Green's function
$G_0$.  This expression turns out to be quite successful. However, in
dense plasmas as well as in a number of materials in solid state physics, 
there is need to improve beyond this simple approximation.  
 
Such an improvement is the $G W^0$ approximation which was studied in 
Ref.~\cite{HB96} for the electron gas at $T=0$
and in Ref.~\cite{WR98} for the solar core plasma. In the latter case,
a considerably broadened quasiparticle and a featureless behaviour
at the plasma frequency was found. No plasmon-like satellite structures
survived the partial self-consistency treatment.

Another  straightforward extension is a
self-consistent solution of Dyson's equation and the screened
interaction while keeping the vertex function in lowest order \cite{HB98}.
This will be termed $G W$ approximation in this paper.
In Ref.~\cite{HB98}, an increase in the quasiparticle bandwidth and a 
featureless satellite structure is found in contradiction to experimental
evidence.
Also, such an approximation leads to a drastic violation of the
f-sum for the inverse dielectric function. The results indicate the
importance of vertex corrections \cite{HA97}. Calculations for a one-dimensional
semiconductor \cite{gbh95} lead to similar conclusions.
Large cancellations between self-energy and vertex corrections
have also been found in Ref.~\cite{R65,GT70,MS89,BTAS97,UBH98}.

Recently, a number of approximation schemes to take into account
vertex corrections have been proposed \cite{td00}. In this work we will apply 
a sequence of approximations described in the following:

Taking the self-energy in the $G W$-approximation, the functional
derivative occurring in the vertex equation~(\ref{eq:vertex}) yields in lowest order of $W_{aa}$
\begin{eqnarray}
\label{eq:dSigmadG_GW}
\frac{\delta \Sigma_a(12)}{\delta G_b(45)}
& = & W^0_{aa}(12) \delta(14) \delta(25) \delta_{ab}~.
\end{eqnarray}
Note that the screened interaction is taken in the one-loop approximation for $\Pi$, the so-called random-phase approximation (RPA).
This term leads to a ladder approximation for the vertex in terms of the 
dynamically screened interaction
\begin{eqnarray}
\Gamma_a(12,3) & = & 
\delta(12) \delta(13) \,+\,
\int\! d(67) W^0_{aa}(12) G_a(16) G_a(72) \Gamma_a(67,3)\,\,\,.
\end{eqnarray}
Already this equation is challenging to solve \cite{NC05}, even in Shindo approximation
\cite{S70}.
As a result, the improved self-energy in this approximation is given
in second order of the dynamically screened interaction by the term studied
in Ref.~\cite{S96},
\begin{eqnarray}
	\Sigma_a^{GW^0\Gamma}(12) & = & G_a(12)W^0(21)\,+\,
   \int\! d(34)\,
   G_a(13)\,W_{aa}^0(23)\,G_a(34)\,W_{aa}^0(41) G_a(42) \,\,\,.
\end{eqnarray}
Also, the improved polarization function is given besides the loop diagram
by an exchange diagram with respect to $W^0$
\begin{eqnarray}
	\Pi^{GG\Gamma}_{aa}(12) & = & 
G_a(12)G_a(21)\,+\,\int\!d(34) G_a(13) G_a(41) W_{aa}^0(43) G_a(24) G_a(32)\,\,\,.
 \,\,\, 
\end{eqnarray}
At this stage the first iteration of Hedin's equations (\ref{eq:Dyson})-(\ref{eq:vertex}) is completed.
We will not go beyond this first iteration, in particular the screened interaction potential $W_{aa}$ is kept on
the level of RPA.


To address the deviations from the quasi-particle picture, the introduction
of the one-particle spectral function $A_a(p,\omega)$ is convenient. With its
help, the spectral representation of the full Green's function is given by
\begin{eqnarray}
\label{eq:spectral_function}
G_a(p,z) & = & \int_{-\infty}^{\infty}\!\frac{\mathrm{d}\omega}{2 \pi}\,
\frac{A_a(p,\omega)}{z-\omega}\,\,\,,
\end{eqnarray} 
$p$ and $z$ denote momentum and energy (frequency) of the particle. The Green's function
$G_a(12)$ is obtained from the Green's function in momentum-frequency representation $G_a(p,z)$ by means of 
Laplace transform.
The quasi-particle approximation itself can be stated as a $\delta$-like form
of the spectral function
\begin{eqnarray}
\label{eq:quasi-particle}
  A_a(p,\omega) & = & 2 \pi \delta\left(\omega-E_a(p)\right) \;\;\;,
\end{eqnarray}
where the energy $E_a(p)$ is obtained as 
\begin{eqnarray}
\label{eq:quasi-particle-energy}
 E_a(p) & = & \frac{p^2}{2 m_a} \,+\,
 \mbox{\rm Re}\,\Sigma_a(p,E_a(p))\,\,\,.
\end{eqnarray}
Note, that the spectral function itself obeys a normalization relation
\begin{eqnarray}
\int_{-\infty}^{\infty} \!\frac{\mathrm{d}\omega}{2 \pi} A_a(p,\omega) & = & 1
\,\,\,. 
\label{eqn:sf_normalization}
\end{eqnarray}
Furthermore, sum rules for the first and second moment of the spectral function
are known \cite{HB96},
\begin{eqnarray}
 \int_{-\infty}^{\infty}\!\mathrm{d}\omega\,\omega\,A_a(p,\omega) & = & E_a^{\rm HF}
\;\;\;, 
\\
 \int_{-\infty}^{\infty}\!\mathrm{d}\omega\,\omega^2\,A_a(p,\omega) & = & 
 \int_{-\infty}^{\infty}\!\mathrm{d}\omega\,\mbox{\rm Im} \Sigma_a^{\rm c}(p,\omega)\,+\,
  \left( E_a(p)^{\rm HF} \right)^2\,\,\,,
\end{eqnarray}
where the index HF refers to the energy in Hartree-Fock approximation
and the index $c$ indicates the use of the correlated self-energy.
From the spectral function, a number of thermodynamic quantities can be obtained,
e.g. the one-particle density \cite{KKER,Vorberger},
\begin{eqnarray}
\label{eq:density_formula}
n_a(\mu_a,\beta) & = & 
\int\!\frac{\mathrm{d}^3p}{(2 \pi)^3} \,\int_{-\infty}^{\infty}\!\frac{\mathrm{d}\omega}{2 \pi}\,
f_a(\omega)\,A_a(p,\omega) \,\,\,, 
\end{eqnarray}
with the distribution function $f_a(\omega)$ of particles of species $a$.

Optical properties, which are under consideration in this work, can be obtained from the well-known
relation between the dielectric function $\epsilon(q,\omega)$ and the 
polarization function Eq.~(\ref{eq:polarization}), i.e.
\begin{eqnarray}
  \label{eq:dielectric_function}
  \epsilon(q,\omega) & = & 1\,-\,\sum_a V_{aa}(q) \Pi_{aa}(q,\omega)\,\,\,,
\end{eqnarray}
with the interaction potential $V_{aa}(q)$.
In particular, the absorption coefficient, which gives the attenuation of electromagnetic radiation
traversing the plasma is given by the imaginary part of the long wavelength limit of the dielectric function as
\begin{equation}
	\alpha(\omega)=\frac{\omega}{c}\mathrm{Im}\,\epsilon(q\to0,\omega),
	\label{eqn:alpha_epsilon}
\end{equation}
relation that holds for wavelengths long against interatomic distances and frequencies high compared to the plasma frequency $\omega_\mathrm{pl}=4(\pi n)^{1/2}$.
For details, we refer to Refs. \cite{Griem,WR98}.

\section{Applications of $GW$ approximation for the self-energy}
\label{sec:app_gw}
The $GW$ approximation has been very successfully used 
for a broad variety of problems in many-particle physics for a long time. 
Here we would like
to mention only a few examples. 

One of the first applications in solid state physics was the calculation of band structures combining density functional theory (DFT)
and the $GW$-method. Northrup et al. \cite{Northrup} showed how this approach improves the band-gap problem of the local density approximation (LDA), which
predicts too small band-gaps for most materials, whereas the use of $GW$ helps to decrease the deviation of the theoretical from the experimental value 
below $0.1\,\mathrm{eV}$ in the case of silicon, compared to about $1\,\mathrm{eV}$ as obtained from pure LDA-DFT. Similar results 
were obtained for many different semi-conductors and insulators \cite{AG98,faleev_prb06}. Optical properties of highly excited semiconductors using
a combined $GW$ and T-Matrix approach have been studied extensively by Schmielau et al., see Ref's \cite{schmielau_pslb00,schepe_pslb98,manzke_cpp01}.
Also in the case of finite systems, the $GW$ approximation has been applied successfully. For the case of the Na$_4$ tetramer, good agreement
between the theoretically predicted photoabsorption and experimental data was obtained \cite{onida_prl95}.

Nuclear and quark matter have been investigated using techniques very similar to the approach presented here. The problem of
dynamical chiral symmetry breaking, which cannot be described in a perturbative calculation of quark propagators, was demonstrated to emerge in
a nonperturbative approach known in quantum field theory as rainbow-ladder approximation \cite{blaschke_pl98,hoell_nucl-th06}.

Fehr et al. \cite{fehr_cpp} and Wierling et al. \cite{WR98} investigated spectral properties of electrons in nonideal plasmas making use of
the $GW$ approximation. Whereas Fehr applied the perturbative $G_0W^0$ approach (c.f. section \ref{sec:GWGamma}) to both equation of state and optical properties, 
Wierling already used a self-consistent $GW^0$ 
approximation which is also used here in the following section.

\section{$GW^0$ approximation for the solar core plasma}
\label{sec:solar}

In the $GW^0$ approximation, the set of Hedin's equation reduces to a self-consistent 
solution of Dyson's equation with the correlated part of the self-energy, 
\begin{eqnarray}
	\label{eqn:sf_sigma}
	A_\mathrm{e}(p,\omega+i\eta) & = &
	\frac{-2\mathrm{Im}\,\Sigma_\mathrm{e}(p,\omega+i\eta)}{\left[%
	\omega-p^2-\mathrm{Re}\,\Sigma_\mathrm{e}(p,\omega+i\eta)\right]^2
	+\left[\mathrm{Im}\,\Sigma_\mathrm{e}(p,\omega+i\eta) \right]^2} \,\,\,,
	\\
	\Sigma_\mathrm{e}(\vec p,\omega+i\eta) & = & 
	-\int\limits_{-\infty}^\infty
	\frac{\mathrm{d}\omega''\,\mathrm{d}\omega'}{(2\pi)^2}
	\int\limits_0^{\infty}\frac{\mathrm{d}^3q}{(2\pi)^3}
	\frac{8\pi}{q^2}\mathrm{Im}\, \epsilon_\mathrm{RPA}^{-1}(q,\omega'+i 0)
	\left[ 1+n_\mathrm{B}(\omega') \right]
	\frac{A_\mathrm{e}(\vec p-\vec q,\omega'')}{\omega+i\eta-\omega''-\omega'} \,\,\,. \nonumber \\  
	\label{eqn:sf_se_coupled_gw} 
\end{eqnarray}
As indicated by $W^0$, the dielectric function $\epsilon(q,\omega)$ is
taken in random phase approximation (RPA) and the Bose function is
given as $n_B(\omega)=\left( {\rm exp}\left(-\omega/T\right)\,-\,1\right)^{-1}$. Here, as in the following, electrons labeled
by $a={\rm e}$ are considered. Note that the Hartree-Fock self-energy, which appears as an additional term in Eq.~(\ref{eqn:sf_se_coupled_gw}) is small for the considered parameters and is henceforth neglected.

\begin{figure}[ht]
	\begin{center}
		\includegraphics[width=.8\textwidth,clip]{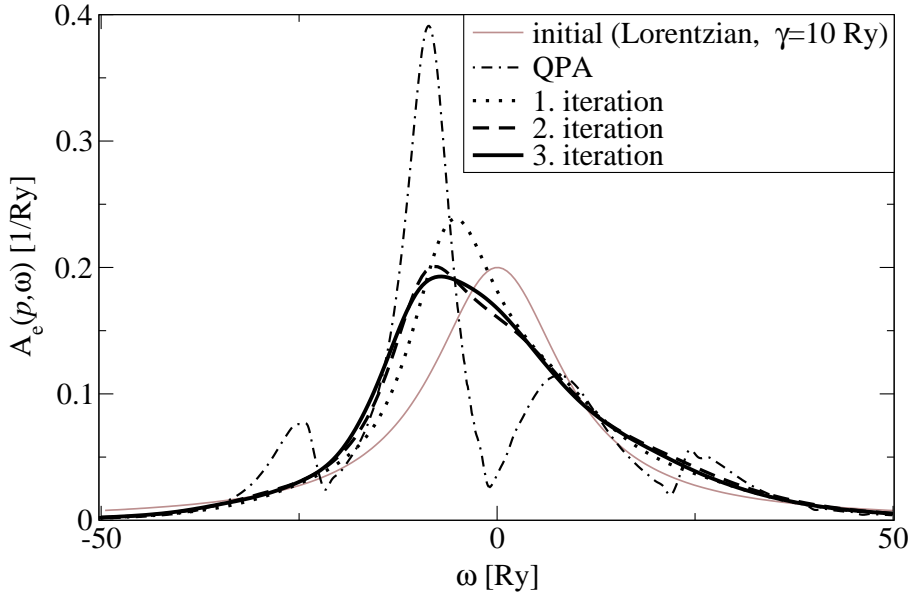} 
		\caption{Electron spectral function in $GW^{0}$ for
solar core conditions ($n_\mathrm{e}=7\cdot10^{24}\,\mathrm{cm}^{-3},
T_\mathrm{e}=100\,\mathrm{Ry}\simeq1360\,\mathrm{eV}$) at momentum $p=0.21 a_\mathrm{B}^{-1}$. Dashed-dotted curve: first
iteration of $GW^0$, (quasiparticle approximation [QPA]). Grey solid curve:
Ansatz for the spectral function in the iterative solution of
$GW^0$. Dotted curve: 1st iteration, dashed: 2nd iteration, solid: 3rd
iteration. The 4th iteration is not distinguishable from the 3rd
iteration.}
		\label{fig:solarcomplete}
	\end{center}
\end{figure}

We give results for a hydrogen plasma using
the conditions at the solar core center. We ignore other ions such as helium etc. in these exploratory calculations.
In Fig.~\ref{fig:solarcomplete}, the self-consistent spectral
function calculated in $GW^0$ approximation, i.e. the solution of the
set of Eqs.~(\ref{eqn:sf_sigma}) and (\ref{eqn:sf_se_coupled_gw}), is shown for a fixed momentum of $p=0.21 a_\mathrm{B}^{-1}$. The 
grey curve is the initial ansatz
for the spectral function, i.e. the input for the r.h.s. of Eq.~(\ref{eqn:sf_se_coupled_gw}). 
It has been chosen of Lorentzian form with a width of $\gamma = 10$ Ry. The
first iteration is given by the dashed-dotted
curve. The peak of the spectral
function is shifted to smaller frequencies and the function is
asymmetrically broadened. The second
and third iteration 
give only minor modifications to the first
iteration, the forth iteration (not shown) does not vary significantly
from the third iteration.

\begin{figure}[ht]
	\begin{center}
 \includegraphics[width=.8\textwidth,clip,angle=0]{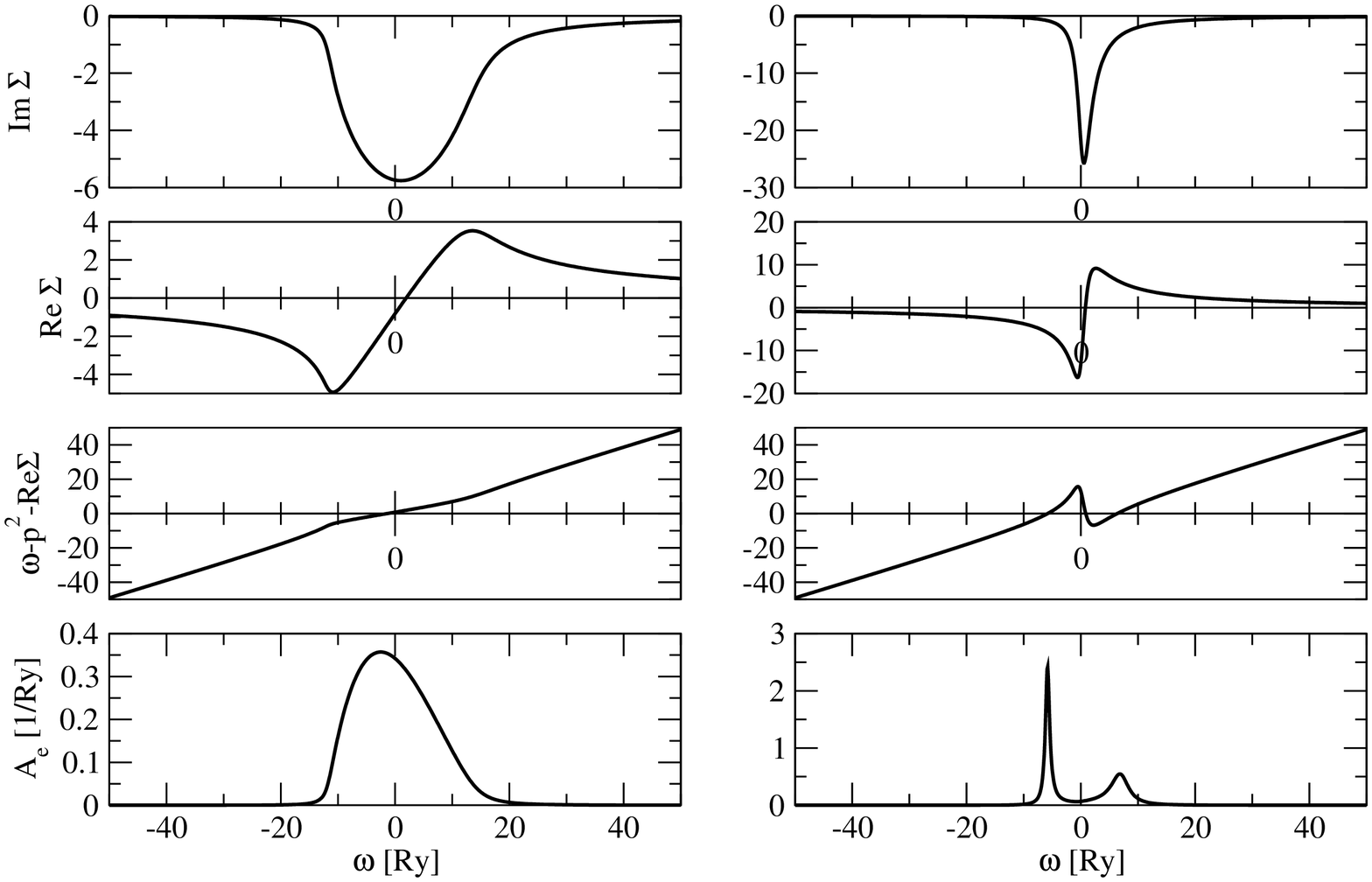}
		\label{fig:se_comp}
		\caption{Self-energy (real and imaginary part), dispersion relation and spectral function
		calculated self-consistently by solving Eq.~(\ref{eqn:se_stat_selfcons}) (left) and quasiparticle
approximation (right), i.e. the r.h.s of
Eq.~(\ref{eqn:se_stat_selfcons}) evaluated with
$\Sigma(\boldsymbol{p},\omega)=0$. Parameters:$n_\mathrm{e}=7\cdot10^{24}\,\mathrm{cm}^{-3},
T_\mathrm{e}=100\,\mathrm{Ry}\simeq1360\,\mathrm{eV}$. The momentum is fixed at $p=0.21\,\mathrm{a_\mathrm{B}^{-1}}$.
			}
	\end{center}
\end{figure}
\begin{figure}[ht]
	\begin{center}
		\includegraphics[width=.8\textwidth,angle=0,clip]{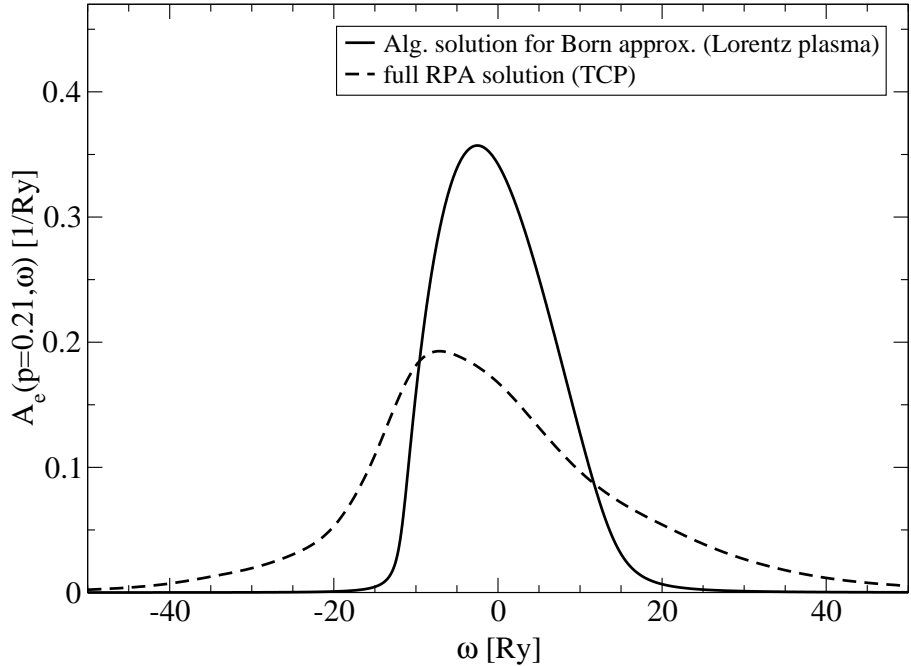}
	\end{center}
	\caption{Self-consistent spectral function computed in full RPA (two component plasma) (dashed curve) and in static Born approximation including only
	electron-ion collisions (solid curve). The neglectance of $e-e$ collisions leads to a smaller width of the spectral function.  Solar core parameters:
	$n=7\cdot 10^{24}\,\mathrm{cm}^{-3}, T=100\,\mathrm{Ry}$.}
	\label{fig:sf_ppa-rpa}
\end{figure}
\begin{figure}[ht]
	\begin{center}
		\includegraphics[width=.95\textwidth,angle=0,clip]{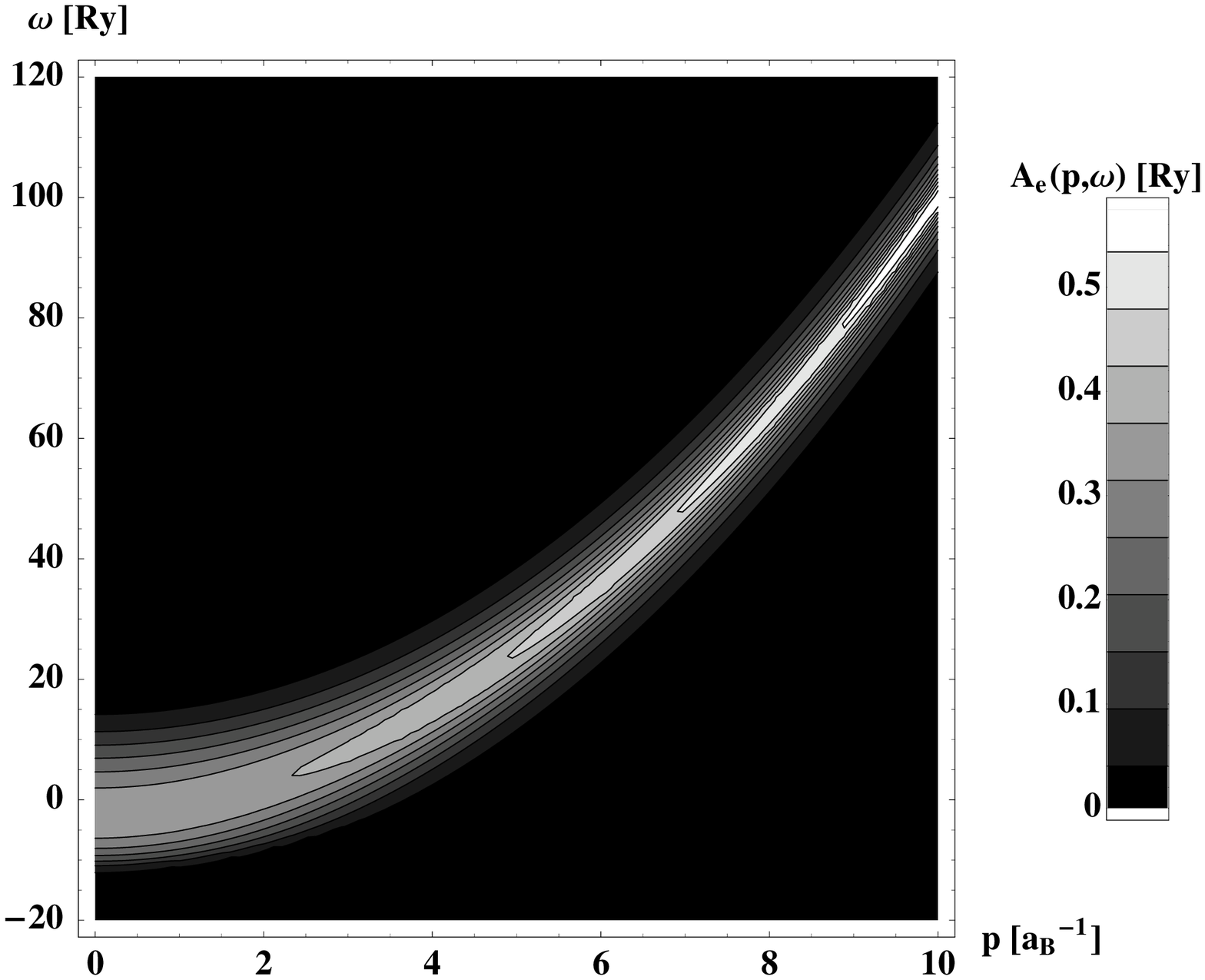}
	\end{center}
	\caption{Contour plot of the self-consistent spectral function with simplified
	self-energy (Eq.~(\ref{eqn:se_stat_selfcons}) at solar core conditions as function of frequency $\omega$ and momentum $p$.
	The spectral function is asymetrically broadened and shifted from the quasi-particle energy $\omega_\mathrm{QP}=p^2$. At high momentum the spectral
	function converges into a sharp quasi-particle resonance at the free-particle energy.}
	\label{fig:contour_sf}
\end{figure}

For comparison, we also show the $G_0 W^0$ approximation, i.e. the first
iteration of Eq.~(\ref{eqn:sf_se_coupled_gw}) starting from a spectral
function of vanishing width (delta-distribution). In this case, there
is no quasiparticle peak at the quasiparticle resonance
$\omega_\mathrm{QP}=p^2$, 
but four satellites and minima appear. The
latter are due to sharp peaks in the response function
$\mathrm{Im}\,\epsilon^{-1}_\mathrm{RPA}(k,\omega)$ and have been observed 
earlier \cite{L67,BYKH}.

As a result of this calculation, we observe, that the self-consistent
calculation leads to a spectral function that is physical easily
understandable, i.e.  it contains a broadened and shifted
quasiparticle resonance. However, the signatures of collective
effects, such as the dynamical screening, which are present in the $G_0 W^0$
approximation (dash-dotted curve in Fig.~\ref{fig:solarcomplete}) vanish
completely in the self-consistent result. In contrast, the
$G_0W^0$ result does not contain a broadened quasiparticle peak but is
completely determined by the behaviour of the response function, which
contains the collective excitations of the plasma (plasmons), as
also shown by Fehr et al., see Ref. \cite{fehr_diss}.

Although convergence is already achieved after 3-4 iterations of Eqs.~(\ref{eqn:sf_sigma}) and (\ref{eqn:sf_se_coupled_gw}), the calculation is too time-consuming to be used in
the computation of physical observables such as equation of state or the polarization function, the latter involving convolution integrals over two
spectral functions in both momentum and frequency domain. Therefore, we make use of some
further approximations. First, we replace the full RPA-like screened interaction
by a one-loop approximation, which takes into account the scattering among particles in Born approximation.
In this case we only consider electron-ion scatterings. 
Ions are treated in adiabatic approximation and the interaction is
mediated by a statically screened potential of Debye-type, i.e.
$V_{\mathrm ei}(q)=8\pi/(q^2+\kappa^2)$, where $\kappa=(8\pi n/T)^{1/2}$ is the inverse Debye screening length. 
This approximation is
justified by the fact that the corresponding absorption cross-section leads to the
nonrelativistic limit of the Bethe-Heitle formula for inverse
bremsstrahlung in the limit of vanishing $\kappa$, as shown in \cite{Fortmann}. 

For small $\kappa$, the main contributions 
to the integral in Eq.~(\ref{eqn:sf_se_coupled_gw}) come from small $q$. Therefore, we can neglect the
shift in the momentum variable in the self-energy on the r.h.s.
As shown in Ref.~\cite{Fortmann}, one then obtains a particularly
simple equation for the self-energy in this approximation, which we will denote by the suffix $GW^0s$,
\begin{equation}
	\Sigma^\mathrm{GW^0s}(\boldsymbol{p},\omega+i\eta)=-8\pi \frac{n_\mathrm{i}}{\kappa}\left[\kappa^2+p^2-\omega-i\eta+
	\Sigma^\mathrm{GW^0s}(\boldsymbol{p},\omega+i\eta)-2i\kappa\sqrt{\omega+i\eta-\Sigma^\mathrm{GW^0s}(\boldsymbol{p},\omega+i\eta)}\right]^{-1}~,
	\label{eqn:se_stat_selfcons}
\end{equation}
which can be solved by standard root-finding algorithms. 
Fig.~2
shows the solution of
Eq.~(\ref{eqn:se_stat_selfcons}) for the same set of parameters as used in the RPA calculation (Fig.~\ref{fig:solarcomplete}). To the left, the self-energy (imaginary
and real part), as well as the dispersion relation, and the spectral
function as obtained from the numerical solution of
Eq.~(\ref{eqn:se_stat_selfcons}) are shown. The spectral function is
centered around the quasiparticle energy. Its overall shape is
given by the imaginary part of the self-energy, while the real part
determines the dispersion. In the self-consistent solution, the
dispersion relation contains only a single root near the quasiparticle energy, while there are three
roots in the quasiparticle approximation (r.h.s of Fig.~2).
The upper and lower root give rise to the two satellites in the
corresponding spectral function, while the central root does not yield
a resonance due to the large imaginary part of $\Sigma^\mathrm{G_0W^0s}$ around
$\omega=1\,\mathrm{Ry}$. The separation of the satellites from the quasiparticle energy is approximately given by the plasma frequency. This is
a general feature, which has been observed in earlier calculations carried out at lower particle densities and temperatures \cite{Fortmann}.

In Fig.~\ref{fig:sf_ppa-rpa} we compare the spectral functions obtained from the two-component RPA and the simplified Born approximation containing only the 
scattering of electrons on fixed ions (Lorentz-Plasma). Since electron-electron collisions are neglected in the latter case, the corresponding 
spectral function is narrower than in the RPA calculation. Furthermore, in the one-loop
	approximation there are no plasmon degrees of freedom which leed to further damping in the RPA calculation. On the other hand, the Lorentz plasma calculation already contains all characteristic
features of the RPA spectral function. Namely the asymmetrical broadening and the shift of the peak towards lower energies show up. These features
become less pronounced at higher momenta, where many-particle effects are expected to be less important. With
increasing momentum, the width of the spectral function decreases, while at the same time the height of the quasiparticle peak increases, as can be seen in Fig.~\ref{fig:contour_sf} . This follows naturally from the
normalization condition for the spectral function Eq.~(\ref{eqn:sf_normalization}). 
In conclusion, we have shown that our simplified Lorentz plasma model for the self-energy leads
physically intuitive one-particle spectral function which includes the electron-ion collisions in a consistent way. 

 \section{Applications of $GW\Gamma$-approximation for the self-energy}
In the last years, substantial effort has been made to go beyond the $GW$-approximation and to include the vertex to some extent. In a pioneering
paper, Takada \cite{T01} demonstrated the feasibility of self-consistent $GW\Gamma$ approximation. He used an ansatz for the vertex-function which fulfils certain sum-rules and
conservation laws. As a result, he showed the subtle cancellation of contributions from self-energy and vertex-corrections to spectral properties of charged
particles, exemplified for the damping of plasmons in Al. 
Very recently, Ziesche \cite{ziesche_annphys06} has reviewed the calculation of direct and exchange contributions (vertex correction) to the on-shell
self-energy of the homogenous electron-gas. 

\begin{figure}[ht]
	\begin{center}
%
%
	\includegraphics[width=.8\textwidth,angle=0,clip]{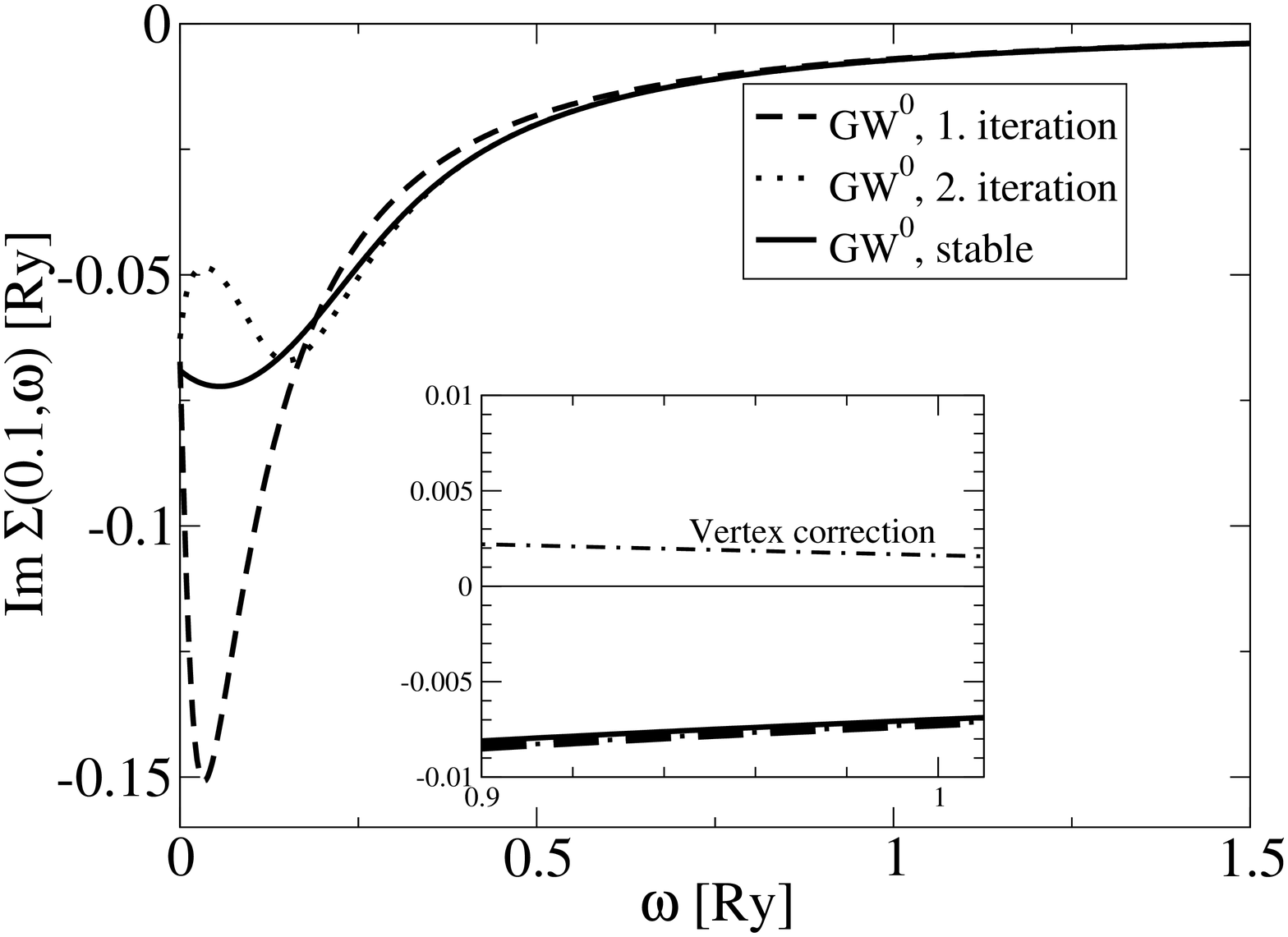}
	\end{center}
	\caption{ Imaginary part of the self-energy as calculated from Eq.~(\ref{eqn:se_stat_selfcons}). The first (dashed) and second (dotted) iteration as well as the convergent result (solid) are shown. Inset: Comparison of $GW$ approximation and lowest order vertex correction (dashed-dotted), cf. Eq.~(\ref{eqn:se_2ndborn}). The exchange diagram gives a correction of at most 20\%.  }
	\label{fig:se_born2_compensation}
\end{figure}

For plasmas, Vorberger et al. \cite{Vorberger} systematically studied contributions to
the equation of state beyond Montroll-Ward, including all exchange terms (vertex corrections). In the same
spirit, we will now investigate the lowest order vertex correction to the self-energy in second order of the interaction, given by the following
diagrammatic expression.

\begin{equation}
	\Sigma^{(2)}(\boldsymbol{p},\omega)=\parbox{2cm}{\includegraphics[width=2cm,clip]{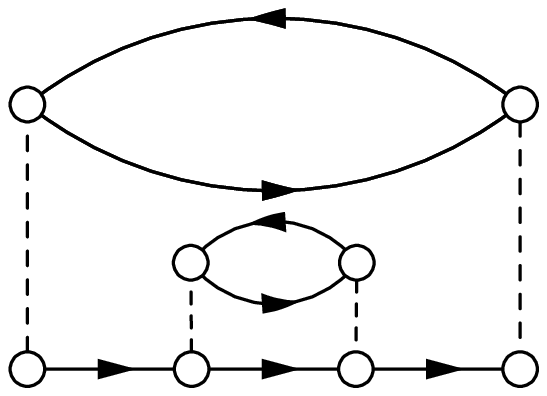}}+\parbox{2cm}{\includegraphics[width=2cm,clip]{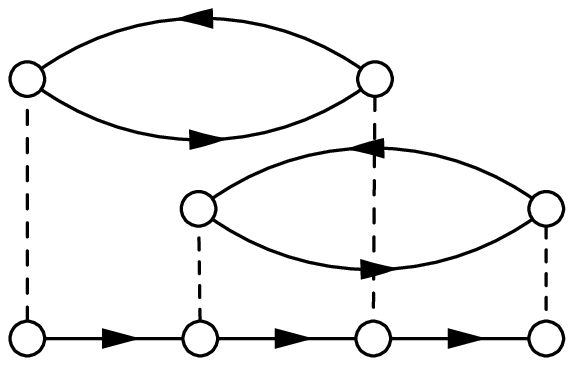}}~.
	\label{eqn:se_2ndborn}
\end{equation}
The second diagram is the lowest order diagram which does not appear in the $GW$ approximation, whereas the first diagram is automatically included.
The base lines are free electron propagators, while the upper loops are
composed of ionic propagators taken in the adiabatic approximation. To ensure convergence, a statically screened potential $V_{\mathrm ei}(q)$ is used.
The numerical evaluation of Eq.~(\ref{eqn:se_2ndborn}) is shown in Fig.~\ref{fig:se_born2_compensation}. The 
dashed
and 
dotted
curves give the first and second iteration respectively of
the $GW^0s$ approximation, the 
solid
curve is the self-consistent
result. For comparison, the first order vertex correction as obtained
from the exchange diagram in Eq.~(\ref{eqn:se_2ndborn}) is shown in
the inset. The vertex term gives at most a 20\% reduction of the
self-energy.

\section{Modification of the absorption coefficient}
\label{sec:absorption}

As an illustrative example, we calculate the impact of the broadened
electron spectral function on the absorption coefficient due to
inverse bremsstrahlung.
The absorption coefficient for radiation in a plasma can be obtained from
the dielectric function $\epsilon(q,\omega)$, Eq.~(\ref{eqn:alpha_epsilon}). 
According to Eq.~(\ref{eq:polarization}), nonideality effects enter via self-energy
corrections of the full Green's functions and by vertex corrections beyond the bare
vertex. Within the $GW$ approximation, only self-energy corrections are considered,
while the $GW\Gamma$ also accounts for vertex terms in the polarization function.

\subsection{Self-energy corrections}
 
\begin{figure}[ht]
	\begin{center}
	\includegraphics[width=.8\textwidth,angle=-0]{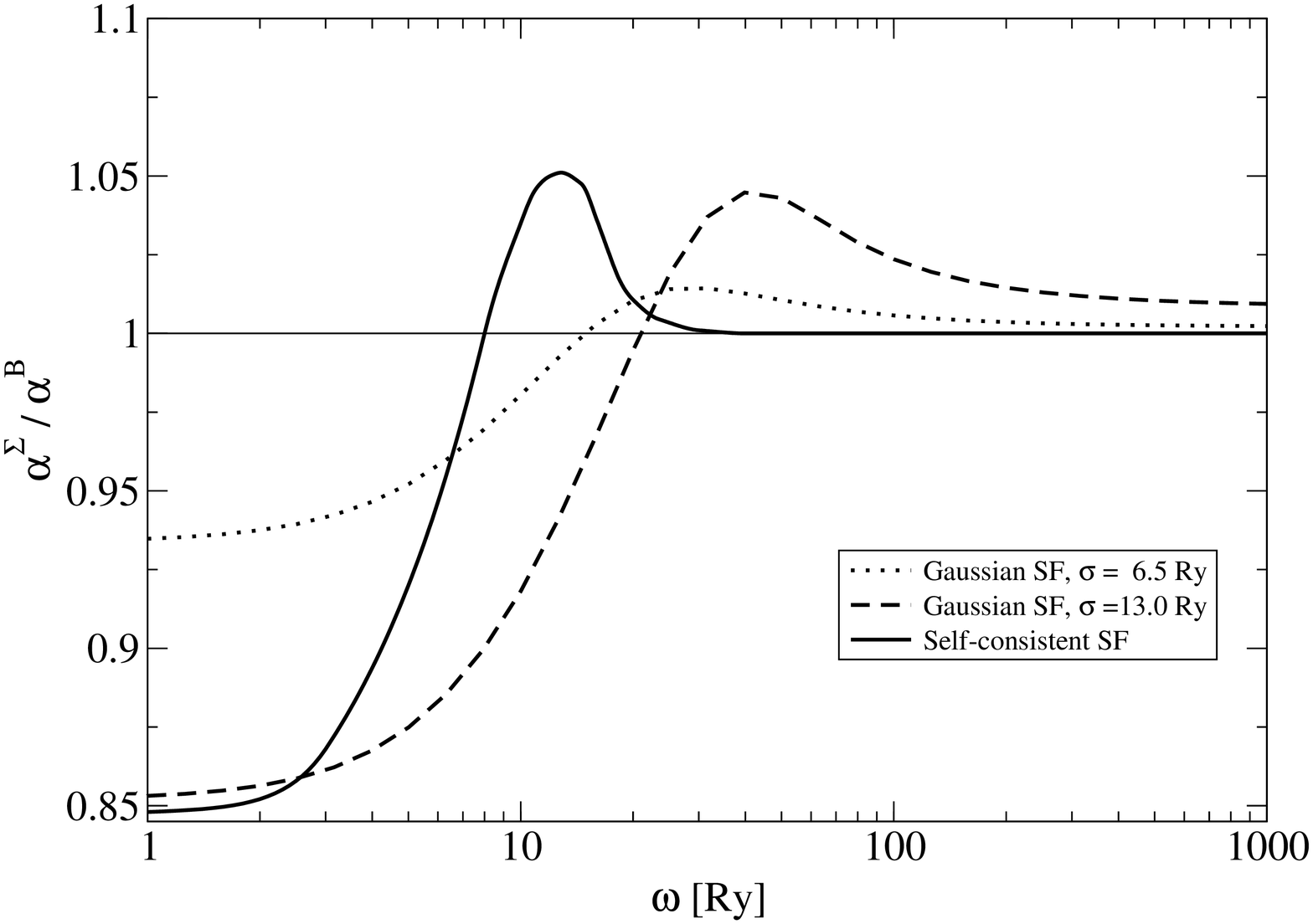}
	\caption{Free-free absorption coefficient calculated
	from broadened electron propagators. The absorption coefficient is
	normalized to the Born approximation. At low frequencies, suppression of
	inverse bremsstrahlung is obtained.
	For high photon energies, the improved result converges to the Born approximation, while around $\omega=10\,\mathrm{Ry}$ enhancement sets in. Also shown:
	$\alpha(\omega)$ calculated with Gaussian spectral functions (dotted and dashed line) of different widths. Here, the convergence into the
	Born result is much slower, since the Gaussian spectral function does not yield the correct quasiparticle limit at high frequencies. Parameters: $n=7\cdot 10^{24}\,\mathrm{cm}^{-3}, T=100\,\mathrm{Ry}$ (solar core).}
	\label{fig:alpha_se}
	\end{center}
\end{figure}

Using the self-consistent spectral function obtained in simplified $GW^0$ (Eq.~(\ref{eqn:se_stat_selfcons})) to calculate the
polarization function, leads us to the result plotted as solid curve in Fig.~\ref{fig:alpha_se}.
We normalize our result to Kramers' formula corrected 
by a Gaunt-factor in Born approximation \cite{karzas67,fortmann_hedp06}, which corresponds to the 
non-relativistic limit of the Bethe-Heitler cross section for inverse Bremsstrahlung \cite{beth:proc.roy.soc34}. In the infrared limit ($\omega\to0$) the Born approximation
shows a logarithmic divergence.

The result for solar core parameters is shown in Fig.~\ref{fig:alpha_se}. 
For small frequencies, a reduction of the free-free absorption
of about 15\% as compared to Born approximation is observed, while at high frequencies our approach converges to the Born approximation.
At intermediate frequencies, we obtain a slight enhancement of our result relative to the Born approximation. These characteristics 
have already been observed in earlier calculations for lower densities and temperatures \cite{Fortmann}.
We compare our results to calculations which use parametrized spectral functions of Gaussian shape in the polarization loop (dashed and dotted curve). 
The width $\sigma$ is given. 
Also in this case, an enhancement is observed which decreases with decreasing width $\sigma$ as can be seen by comparing the dotted ($\sigma=6.5\,\mathrm{Ry}$) and
the dashed curve ($\sigma=13\,\mathrm{Ry}$). The value $\sigma=6.5\,\mathrm{Ry}$ gives a Gaussian spectral function of similar shape as the self-consistent
calculation at small momenta. As shown in Fig.~\ref{fig:contour_sf}, the self-consistently calculated spectral function converges to a quasi-particle resonance
at large momenta, i.e. large frequencies. This behaviour leads to the faster convergence of the absorption coefficient to the Born result as compared to the 
calculation using Gaussians with frequency and momentum independent widths. 
Remember that the Born result for the absorption coefficient is obtained by inserting delta-like spectral functions
in the polarization function Eq.~(\ref{eq:polarization}).

\subsection{Vertex corrections}
\label{sse:vertex_corrections}

We have shown that the account of a broadened one-particle spectral function leads to a
suppression of the infrared behaviour of the inverse bremsstrahlung
spectrum. 
However, the calculations have been carried out on the level
of $GW^0$-approximation. Vertex corrections contribute to the polarization
function in the same order with respect to the statically screened potential.
Moreover, these vertex corrections tend to cancel the self-energy correction to some extent.
In order to account for both types of corrections on the same footing, 
we focus on contributions from the vertex equation, Eq.~(\ref{eq:vertex}), 
up to first order in the screened interaction
in the polarization, Eq.~(\ref{eq:polarization}),
\begin{equation}
	\Pi(\boldsymbol{q},\omega)=\parbox{2cm}{\includegraphics[width=2cm,clip]{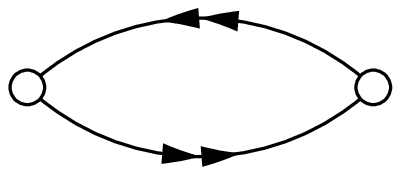}}+\parbox{2cm}{\includegraphics[width=2cm,clip]{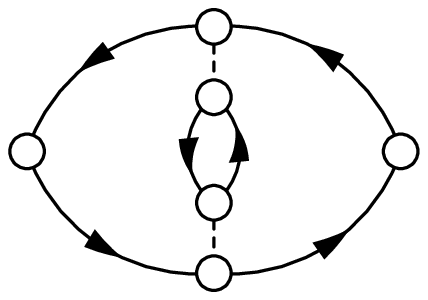}}~.
	\label{eqn:pol_2ndborn}
\end{equation}
\begin{figure}[ht]
	\begin{center}
	\includegraphics[width=.8\textwidth,angle=0]{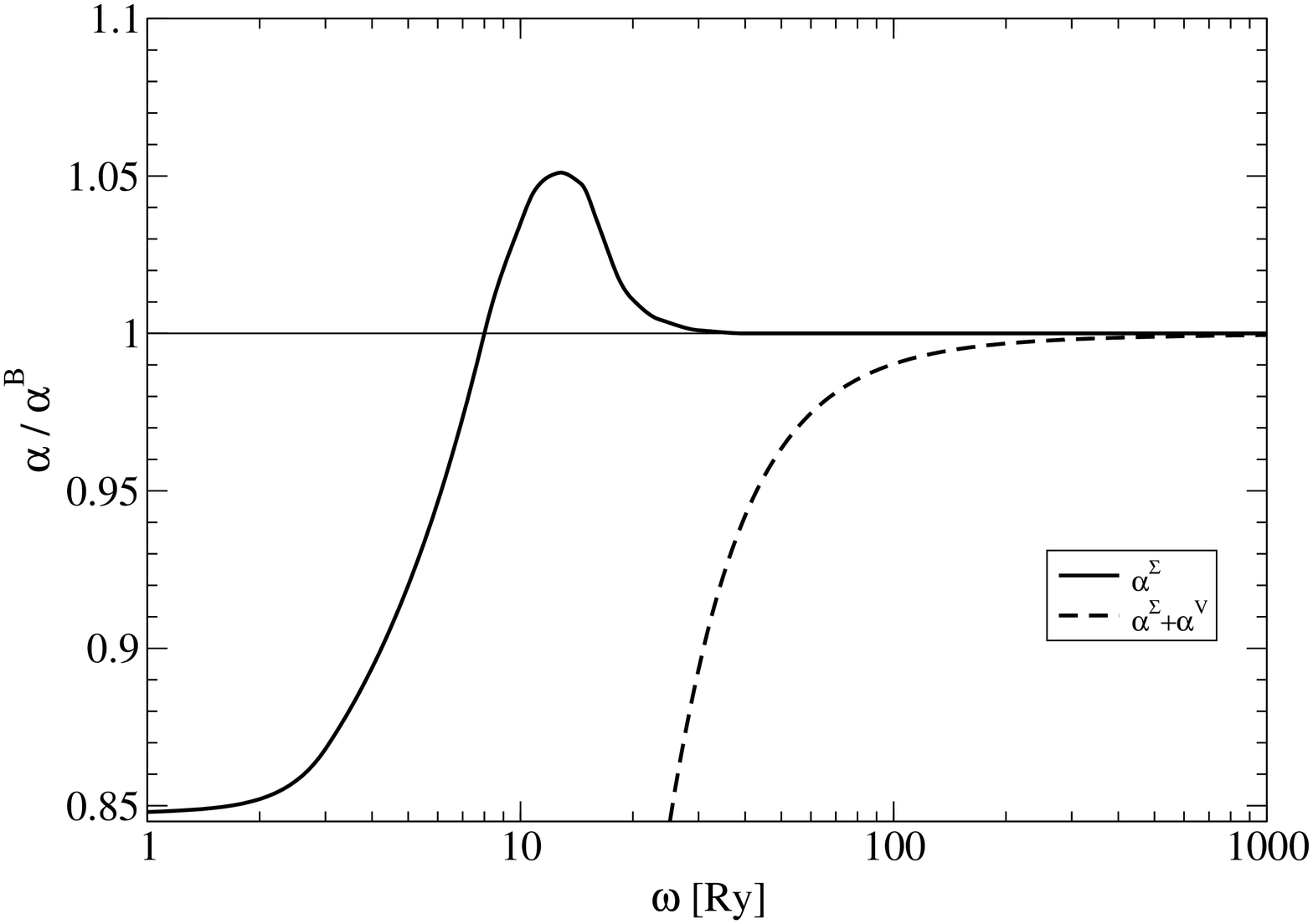}
	\end{center}
\caption{ Absorption coefficient $\alpha(\omega)$ for solar core conditions as function of the
photon energy $\omega$. $\alpha(\omega)$ is calculated from
Eq.~(\ref{eq:polarization}) with broadened electron propagators using
the self-energy obtained from Eq.~(\ref{eqn:se_stat_selfcons}) (solid
curve). Additionally, the first vertex correction is calculated,
cf. Eq.~(\ref{eqn:se_2ndborn}). The sum of both terms is given by the
dashed curve. The vertex correction dominates the behaviour of the absorption coefficient. In particular,
the enhancement at $\omega\simeq 10\,\mathrm{Ry}$ present in the pure self-energy calculation
vanishes completely.} \label{fig:alpha_sum}
\end{figure}

In Fig.~\ref{fig:alpha_sum} we compare the suppression of the absorption coefficient obtained from the inclusion of
self-energy only (%
solid 
curve) and with included vertex correction in the polarization loop (%
dashed 
curve), c.f. Eq.~(\ref{eqn:pol_2ndborn}). At the present parameters, the vertex correction
is the most dominant term. Over the whole range of frequencies considered here, we obtain a suppression
of the absorption coefficient as compared to the Born approximation. Also, the enhancement of absorption at
frequencies around $\omega=10\,\mathrm{Ry}$, which was observed in the calculation using only self-energy corrections (solid curve),
vanishes if the vertex is taken into account (dashed curve). This shows the importance of vertex corrections.

Here, the vertex correction is only taken in lowest order of
the interaction and density, while the self-energy based result contains a summation to all orders of density. In order to
compare self-energy and vertex contributions to the modification of the absorption spectrum in a fully consistent way, 
one would have to go beyond the
perturbative calculation presented here and solve for the vertex equation Eq.~(\ref{eq:vertex}) using at least free-particle
propagators. This task goes beyond the scope of this paper, where only the lowest order corrections are to be studied. Finally,
we remark, that in the mentioned earlier calculations presented in Ref.~\cite{Fortmann}, the vertex correction modifies the
self-energy result only to a minor extent, i.e. the enhancement at intermediate frequencies is reduced by 40\%. This is due to the
lower density ($10^{19}\,\mathrm{cm^{-3}}$) used in that work.
In all calculations, the high frequency behaviour converges nicely to the Born result. 

For the infrared part of the spectrum, the behaviour of the absorption coefficient needs further considerations, going beyond the
present work. As shown in Fig.~\ref{fig:alpha_sum}, the self-energy correction in the polarization function reduces the absorption coefficient by a constant factor of about 15\% in
the low frequency limit, but 
does not regularize the infrared divergence of the Born approximation.
The vertex term, on the other hand, induces a large suppression so that in the low frequency limit higher orders of
the vertex correction have to be considered. 
Furthermore, it is well-known, that below the plasma frequency ($7\,\mathrm{Ry}$ for the solar core parameters) dynamic screening plays an important r\^ole. 
This needs a further summation of diagrams going beyond the present scheme.
The impact of dynamical screening on
bremsstrahlung emission has been discussed by Ter-Mikaelyan in Ref.~\cite{TerMikaelyan}.

\section{Conclusions}
\label{sec:conclude}
A systematic treatment of optical properties in non-ideal plasmas is
possible in the framework of Green's function methods. Corrections beyond
the quasi-particle pictures can be generated using Hedin's equations. 
However, a consistent solution of Hedin's equations is a formidable task. 
Here, we
considered a simplified set of equations. Notably the $GW_0$-approximation, 
mainly used in solid state physics, 
has been shown to lead to sensible results also in the field of plasma physics. 
Results for the self-energy and the spectral function are presented. Plasmon-like structures, present in
perturbative calculations as performed in Ref. \cite{fehr_diss} vanish completely. An asymmetrically broadened and shifted
spectral function is obtained. Furthermore, it is
shown, that a simplified model, where the dynamically screened 
interaction is approximated by static screening leads to similar results for both
self-energy and spectral function as compared to the full RPA results. 
Deviations can be understood as a consequence of neglecting collisions among particles of equal species.

The simplified $GW^0$ approximation for plasmas is now available for a broad range of parameters, i.e. density and temperature. Also,
the correct asymptotic behaviour, i.e. convergence to a delta-like quasiparticle resonance at large momenta, is obtained. 
In this paper we used parameters corresponding to the solar core, whereas
Ref. \cite{Fortmann} contains similar results for lower density and temperature.

Vertex corrections to the self-energy have been studied on a perturbative level. So far, only the high frequency
behaviour of the exchange diagram second order in the screened interaction was evaluated. It leads to
20\% reduction of the self-energy with respect to the second order of $GW^0$.

The availability of the one-particle spectral function for any momentum and frequency makes it possible to use 
it in calculations of further physical observables as the equation of state and the optical properties. Here we focused on
the absorption of electromagnetic radiation due to free-free transitions, which is obtained from the polarization function.
Insertion of the broadened particle propagators in the one-loop approximation for $\Pi$ 
leads to a significant modification of the
absorption spectrum at low frequencies. Besides the suppression in the infrared, enhancement of absorption is
observed at intermediate frequencies. 

Since the one-loop approximation using full propagators is an inconsistent approximation with
regards to conservation laws, such as Ward identities, we investigated the vertex-correction inside the
polarization loop in lowest order. It turned out, that the vertex correction is by far the most important 
correction to Born approximation, i.e. the self-energy effects contained in the full spectral functions, are dominated
by the vertex correction. However, this is not a general feature as becomes clear if comparing to calculations for other parameters in
Ref. \cite{Fortmann}. On the other hand, a consistent comparison of both self-energy and vertex corrections on the same level of approximation necessitates the
summation of vertex-corrections by solution of the Bethe-Salpeter equation (\ref{eq:vertex}). This task will be accomplished in the future.

\section*{Acknowledgements}
This article was supported by the DFG within the Sonderforschungsbereich 
652 'Starke Korrelationen und kollektive Ph\"anomene im Strahlungsfeld: 
Coulombsysteme, Cluster und Partikel.' A.W. would like to thank the Center of
Atomic and Molecular Technologies of Osaka University for its hospitality.
C.F. acknowledges stimulating discussion with J. Vorberger and W.D. Kraeft and thanks
R. Zimmermann for many helpful advice.


\begin{thebibliography}{10}
\bibitem{A} In plasma physics, the use of $V^s$ instead of $W$ is more
            common.
\end{thebibliography}
\end{document}